\documentclass[aps,amssymb,prl,twocolumn,floatfix]{revtex4}
\usepackage[dvips]{graphicx}

\def\bn{\begin{itemize}}
\def\en{\end{itemize}}

\begin{document}
\bibliographystyle{apsrev}
  
\title{Energy Barrier Scalings in Driven Systems}
\author{Craig~E.~Maloney$^{(1,2)}$}
\author{Daniel~J.~Lacks$^{(3)}$}
\affiliation{
    $^{(1)}$ Department of Physics, University of California, Santa Barbara, California 93106, USA}
\affiliation{$^{(2)}$ Lawrence Livermore National Lab - CMS/MSTD, Livermore, California 94550, USA}
\affiliation{$^{(3)}$ Department of Chemical Engineering, Case Western Reserve University, Cleveland, Ohio 44106, USA}
\date{\today}

\begin{abstract}
Energy landscape mappings are performed for two different molecular systems under mechanical loads.  
Barrier heights are observed to scale as $\Delta U\sim\delta^{3/2}$, where $\delta$ is a residual load.  
Catastrophe theory predicts that this scaling should arise for vanishing $\delta$, however, this region is irrelevant in physical processes at finite temperature because thermal fluctuations cause the system to cross over the barrier before reaching the small $\delta$ regime.
Surprisingly, we find that the $\Delta U\sim\delta^{3/2}$ scaling is valid far beyond the vanishing $\delta$ regime described by catastrophe theory.  
This scaling will therefore be relevant at finite temperatures, and can be the basis for corrections to standard rate theoretic approaches.
\end{abstract}
\maketitle
In a broad range of condensed matter systems, one is interested in the question of how some material responds to an external mechanical load. 
External loads cause liquids to flow, in Newtonian or various types of non-Newtonian flows.  
Glassy materials, composed of polymers, metals, or ceramics, can deform under mechanical loads, and the nature of the response to loads often dictates the choice of material in various industrial applications.  
In biological systems, the response of proteins to external loads governs aspects of cell adhesion and muscle function.

The nature of all of these responses depends on both the temperature and loading rate.  
As described by Eyring~\cite{Eyring36}, mechanical loading lowers energy barriers, thus facilitating progress over the barrier by random thermal fluctuations.  
The Eyring model approximates the loading dependence of the barrier height as linear.
The Eyring model, with this linear barrier height dependence on load, has been used over a large fraction of the last century to describe the response of a wide range of systems~\cite{LarsonBook} and underlies modern approaches to biophysical rupture processes~\cite{Bell78,EvansR97,Craig2004}, sheared glasses~\cite{Argon79,RottlerR03}, \emph{etc.}

The linear dependence will always correctly describe small changes in the barrier height, since it is simply the first term in the Taylor expansion of the barrier height as a function of load.  
It is thus appropriate when the barrier height changes only slightly before the system escapes the local energy minimum.  
This situation occurs at higher temperatures; for example, Newtonian flow is obtained in the Eyring model in the limit where the system experiences only small changes in the barrier height before thermally escaping the energy minimum.

As the temperature decreases, larger changes in the barrier height occur before the system escapes the energy minimum (giving rise to, for example, non-Newtonian flow).  
In this regime, the linear dependence is not necessarily appropriate, and can lead to inaccurate modeling.  
For example, Li and Makarov~\cite{LiM03} have shown that there is a nonlinear barrier height dependence in stretched proteins, and that the assumption of a linear dependence in the analysis of experimental results leads to inaccurate conclusions.

The present investigation addresses this load dependence of the barrier height.  
The analysis is based on the energy landscape formalism~\cite{StillingerW82}, which considers dynamics to be the sum of vibrational-like motion within energy minima and transitions between energy minima.  
Barriers are associated with saddle points that connect adjacent energy minima.

\begin{figure}
    \hfill
    \includegraphics[width=.2\textwidth]{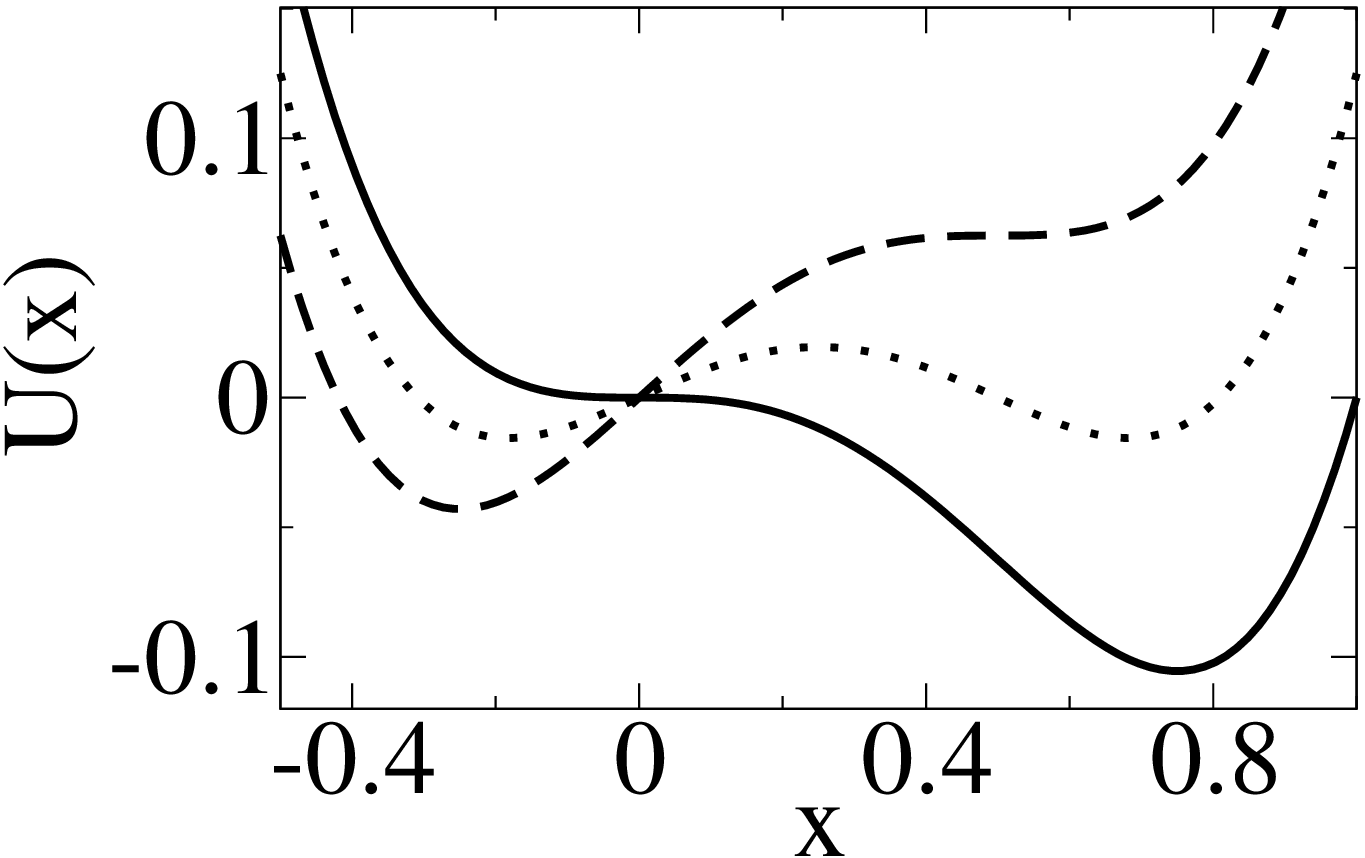}
    \hfill
    \includegraphics[width=.2\textwidth]{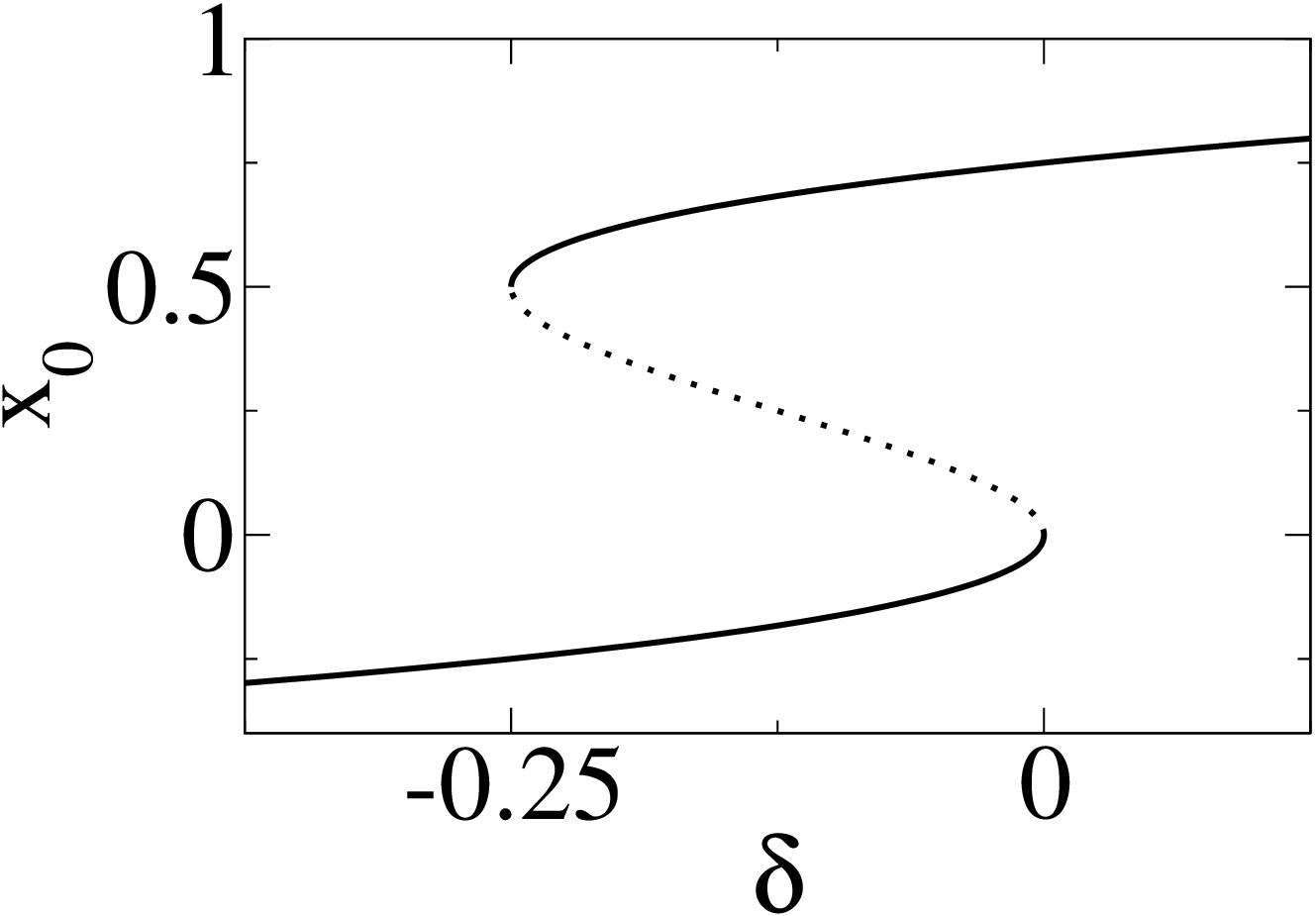}
    \hfill
  \caption{
    Left: Energy from equation~\ref{eq:fourthOrder} for various $\delta$. $\delta=0$: solid, $\delta=-1/8$: dotted, $\delta=-1/4$: dashed.
    Right: Bifurcation diagram indicating the locations of the extrema as a function $\delta$. Minima: solid, Barrier: dotted.
    At $\delta=0$ the minimum of interest collides with a barrier.
    At $\delta=\delta_{min}=-1/4$, the barrier collides with a distant minimum and ceases to exist; it makes no sense to discuss quantities such as $\Delta U$ for $\delta<\delta_{min}$
  }
  \label{fig:fourthOrderCartoon}
\end{figure}

In molecular systems, the energy is a smooth function of the internal degrees of freedom plus a control parameter (\emph{e.g.} the controlled stress or strain).
\footnote{The smoothness underlies the results we present below, and differences would arise in discontinuous models~\cite{EvansR97,HummerS03}.}
As the control parameter is varied, any minimum in the landscape will flatten out in some direction as the minimum collides with a first order saddle point collides ~\cite{MalandroL98,MaloneyL04b,Lacks05} (\emph{c.f.} figure~\ref{fig:fourthOrderCartoon}).
This type of externally induced topological change in a function is known as a fold catastrophe~\cite{ArnoldBook}.

It has long been appreciated that these fold catastrophes induce universal scalings of particular features of the surface in the limit where the minimum and saddle point are infinitesimally close together ~\cite{ArnoldBook}.  In this limit, the function has the lowest order Taylor expansion 
\footnote{ The irrelevant directions (orthogonal to the reaction coordinate) on the PEL can be taken into account in a straightforward way, entering quadratically. They do not change the scalings, and we do not discuss them further.}
\begin{equation}
  U=-Ax^3-Bx\delta 
  \label{eq:fold}
\end{equation}
because the first order term (in the internal degrees of freedom) is zero at a minimum or saddle point, and the second order term (projected along the direction which connects the minimum and barrier) is zero at the point where the minimum and saddle point collide. In equation~\ref{eq:fold}, $x$ is the projection of the system's coordinates onto the zero curvature direction, $\delta$ is the control parameter's distance away from the singularity, and $A$ and $B$ are positive constants. 
To obtain the scaling laws, we note that the minimum and saddle point, $x_{-}$ and $x_{+}$, are the points where the energy derivative vanishes, the curvature along the reaction coordinate at the minimum and saddle point correspond to the second partial derivatives of the function, $\lambda_-=\left.\frac{\partial^2 U}{\partial x^2}\right|_{x_{-}}$, $\lambda_+=\left.\frac{\partial^2 U}{\partial x^2}\right|_{x_{+}}$, and
the height of the barrier, $\Delta U$, corresponds to $\Delta U=U(x_{+})-U(x_{-})$.
This analysis leads to the following scaling relations:
\begin{eqnarray}
  -x_{-}=x_{+} & = & \sqrt{\frac{-B \delta}{3A}}\label{eq:s0}\\
  \lambda_-=-\lambda_+ & = & 6A\sqrt{\frac{-B \delta}{3A}}  \label{eq:curve}\\
  \Delta U & = & 2A\left(\frac{-B \delta}{3A}\right)^{3/2}\label{eq:deltaU}
\end{eqnarray}
The fold ratio, $r_f\doteq6 \Delta U/(2\lambda_- x_-^2)$, is unity when all three of these scaling relations are valid. ~\cite{Wales01}  
These arguments appeal only to $\delta$'s role as a control parameter and are equally valid when $\delta$ represents, \emph{e.g.}, an imposed strain or stress. Recently, the consequences of these scalings have been discussed quantitatively in the context of phenomenological models~\cite{DudkoFKU03,Bogdan2004,ChenC05}.
Fold ratios in incipient catastrophes have been measured in molecular simulations~\cite{Wales01}, but the individual scaling relations have not previously been addressed in externally driven molecular level simulations.

These scaling relations must be obtained in the limit $\delta\rightarrow 0$, but at finite temperature, thermal fluctuations cause the barrier to be crossed before this vanishing $\delta$ regime is being reached.  
Little attention has been paid to the accuracy of these scaling relations for the physically meaningful finite $\delta$ regime.  
While the fold ratio has been shown to deviate significantly from unity at finite $\delta$,~\cite{Wales01} the accuracy of the scaling relations for the individual quantities has not previously been addressed.

Clues to how the scaling breaks down at finite $|\delta |$ are obtained from a 1+1D energy function, $U(x,\delta)$, based on arguments similar to those given in~\cite{MaloneyL04b}.
The only requirement we make of $U$ is that the coupling to the control parameter is bi-linear over the region of interest: $U(x,\delta)=U_0 (x) - B_0 x\delta$.
Demanding that $\left(\partial U/\partial x\right)$ remain zero as we change $\delta$, requires that
\begin{eqnarray}
  \frac{d}{d\delta}\left(\frac{\partial U}{\partial x}\right) & = &\frac{\partial^2 U}{\partial x\partial \delta}+\left(\frac{dx_0}{d\delta}\right) \left.\frac{\partial^2 U}{\partial x^2}\right|_{x_0} \\
  \Rightarrow \frac{dx_0}{d\delta} & =& \frac{B_0}{\lambda_0} \label{eq:curveDist}
\end{eqnarray}
where $x_0$ is a stationary point.
The energy of a stationary point then changes according to
\begin{equation}\frac{dU}{d\delta}=\left(\frac{dx_0}{d\delta}\right)\left.\frac{\partial U}{\partial x}\right|_{x_0}+\frac{\partial U}{\partial \delta}=\frac{\partial U}{\partial \delta}=-B_0 x_0.
  \label{eq:s0DeltaU}
\end{equation}
where the second equality follows from mechanical equilibrium.
The fold scalings obey the above relations between the energy, position and curvature of a stationary point, but these latter relations are more general because they are based on much weaker assumptions about the form of the energy function than the cubic form used to obtain the fold scalings.
Since the energy barrier is obtained after two integrations of the inverse curvature, we anticipate that as the load is backed away from the catastrophe deviations from the scaling relations should occur first for the curvature then for the position and finally for the energy; \emph{i.e.}, the barrier height scaling should be more accurate at finite $\delta$ than the other scaling relations.

We first test this hypothesis regarding the relative accuracy of the barrier height scaling on a simple, analytically solvable, model.  This simple model includes the next order term in the Taylor expansion for the energy, giving a 4-th order polynomial.
\begin{equation}
  U=-x^3-x\delta+x^4 
  \label{eq:fourthOrder}
\end{equation}
The landscape for this energy function for various $\delta$ is shown in figure~\ref{fig:fourthOrderCartoon}~\footnote{Prefactors on each of the three terms may be absorbed into a redefinition of length, energy, and $\delta$.}
.
As $\delta$ goes to zero, the (left hand) minimum and the energy barrier join together at $x=0$.
As $\delta$ is backed away from zero, the minimum and barrier move apart.
As $\delta$ is backed further away from zero, the barrier eventually collides with some \emph{other} minimum at $\delta=-1/4$, rendering the quantities $\Delta U$, $\lambda_+$, $x_{+}$ undefined.
The results for this simple model are consistent with our expectation that the scaling of the barrier height will be more accurate than the scaling of the curvature: at the largest possible values of $\delta$, the barrier height scaling remains accurate to within 10\%, 
while the scaling for $\lambda_-$ is in error by 100\%, and the scaling prediction for $\lambda_+$ is infinite in error ($\lambda_+$ vanishes altogether).
Analogous results are obtained with a negative fourth order term, in which case, the maximal $|\delta |$ occurs when the minimum collides with the other barrier (in this case $\lambda_{+}$ is at 50\% of the value from the scaling relation, and $\lambda_{-}$ vanishes altogether).  
Similar results are obtained for other simple, analytically solvable, models.

\begin{figure}
  \includegraphics[width=.48\textwidth]{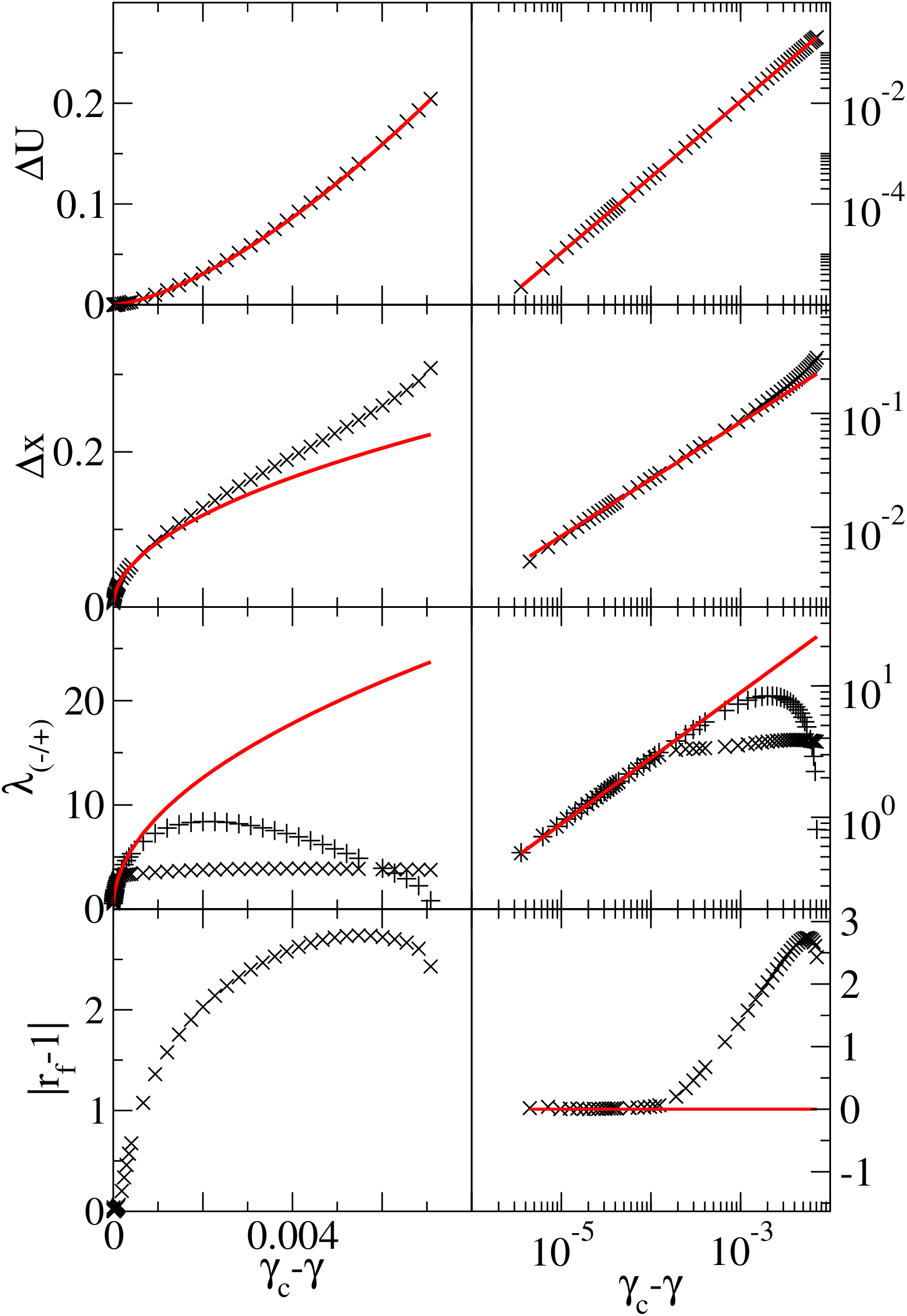}
  \caption{From top to bottom:$\Delta U$, $\Delta x$, $\lambda_{-}$ and $\lambda_{+}$ (plus symbols), and $|r_f -1|$ upon approach to a typical catastrophe.  The solid (red) lines are the theoretical scaling predictions (equations,~\ref{eq:s0},~\ref{eq:curve}, and~\ref{eq:deltaU}) with prefactors determined by a fit to the smallest two decades of $\gamma_c -\gamma$ shown.  $\gamma_c$ is determined via optimization of $\Delta U$ to the $(\gamma_c -\gamma)^{3/2}$ form.}
  \label{fig:lj}
\end{figure}

We have tested these ideas in simulations of realistic atomistic models by tracking local minima and saddle points as a control parameter is varied.  To check the generality of our arguments, we investigate two very different systems (a model glass, and a model protein), and consider both    strain and stress as control parameters.  
The model glass simulations use the Stillinger-Weber 80:20 mixture~\cite{StillingerW84b}, and include 500 particles in an orthorhombic simulation cell with periodic boundary conditions (the Stillinger-Weber functional form, which is similar to the Lennard-Jones potential, is used because the continuous derivatives at a finite cutoff are necessary to analyze the energy landscape at the required precision).  
The model protein is the Thirumalai model~\cite{GuoTH92}, which consists of 46 sites that interact with bond stretching, angle bending, torsion, and nonbonded (Lennard-Jones) interactions.
A minimum is tracked by repeatedly minimizing the energy as the control parameter is varied in small increments.
Similarly, a saddle point is tracked by repeatedly minimizing the sum of the squares of the forces as the control parameter is varied in small increments (the saddle point is found initially by searching halfway along the vector that connects two minima).
All numerical minimizations are performed using a variable metric algorithm, and the eigenvalues and eigenvectors of the system are computed using a standard $QL$ reduction algorithm~\cite{numericalRecipes}.

Results for the glass, with shear strain as the control parameter, are shown in figure~\ref{fig:lj}.
Since the system is multidimensional, it will have many eigenvalues at the minimum, and we take $\lambda_{-}$ to be the smallest of these.
The magnitude of the single negative eigenvalue at the barrier is denoted as $\lambda_{+}$.
We choose to use $\lambda_{-}$ to compute the fold ratio.

In the small $\delta$ limit all of the scaling relations (equations~\ref{eq:s0},~\ref{eq:curve}, and~\ref{eq:deltaU}) are accurate, and the fold ratio is unity.  This behavior is of course expected, because the catastrophe is of the fold type.  In terms of the large $\delta$ behavior, the results are fully consistent with the ideas based on the arguments above, namely:
(i) the barrier eigenvalue goes to zero at about $\gamma_c - \gamma \sim .007$ in a collision with some \emph{other} minimum.
(ii) the accuracy of the scaling relation for the curvature (~\ref{eq:curve}) is quite poor in comparison with the accuracy of the scaling relation for the barrier height (~\ref{eq:deltaU}), with the barrier height scaling being a reasonable approximation over the entire interval up to $\gamma_c - \gamma \sim .007$.
\footnote{
  The sharp transition in $\lambda_{-}$ (green diamonds) in figure~\ref{fig:lj} can be understood in terms of multidimensional hybridization effects.
  Above $\gamma_c - \gamma \sim 10^{-4}$, other eigenmodes have eigenvalues that are lower than the curvature along the reaction coordinate, and it is only below $\gamma_c - \gamma \sim 10^{-4}$ that the reaction coordinate becomes a \emph{bona fide} eigenmode. 
  The eigenmode at the \emph{barrier}, on the other hand, is always reasonably well aligned along the reaction coordinate.
}

\begin{figure}
  \includegraphics[width=.48\textwidth]{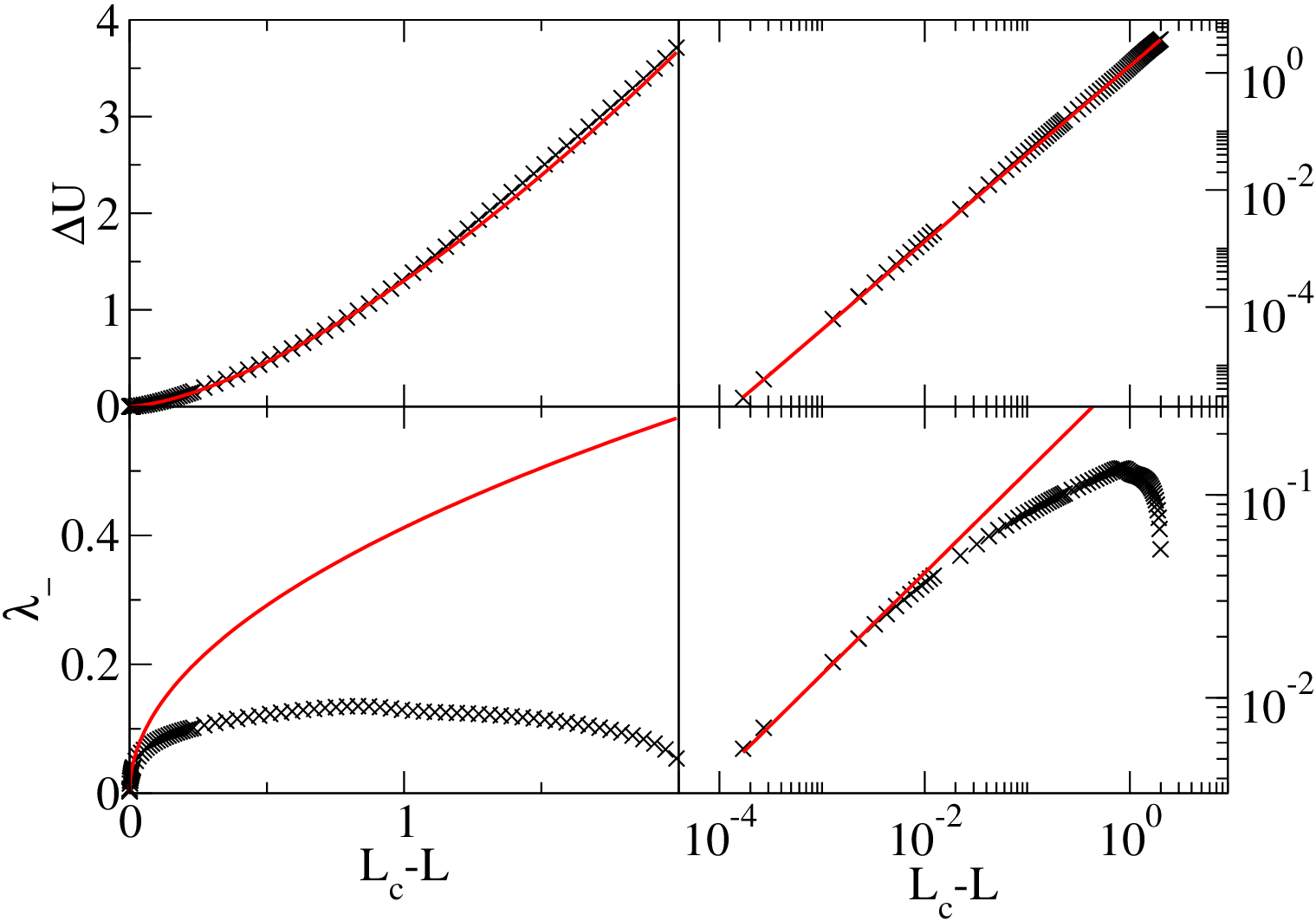}
  \caption{Top: $\Delta U$ and Bottom: $\lambda_-$ for the protein model in a length controlled mode as functions of $L_c-L$ where $L$ is end to end length and $L_c$ is the value of length at which the minimum and barrier collide.}
  \label{fig:proteinLength}
\end{figure}

For the protein model, the end to end distance of the protein, $L$, can be taken as the control parameter. 
As shown in figure~\ref{fig:proteinLength}, both increasing and decreasing $L$ causes the lowest curvature at the minimum to go to zero, indicating the onset of catastrophes.
As expected, the scaling relations are accurate at small values of $\delta$, where $\delta$, is $L_c-L$ and $L_c$ is the length at which the minimum and barrier collide.  At large $\delta$, the scaling relation for the curvature becomes poor while the scaling relation for the barrier height is reasonably accurate over the entire range of $\delta$.
In contrast to the Lennard-Jones case, it is the minimum which disappears in a collision with an extraneous barrier with $\lambda_{-}$ going to zero at the maximal $L_c -L$, but, in both cases, the barrier scalings are found to be superior to the scalings of either of the curvatures.

\begin{figure}
  \includegraphics[width=.48\textwidth]{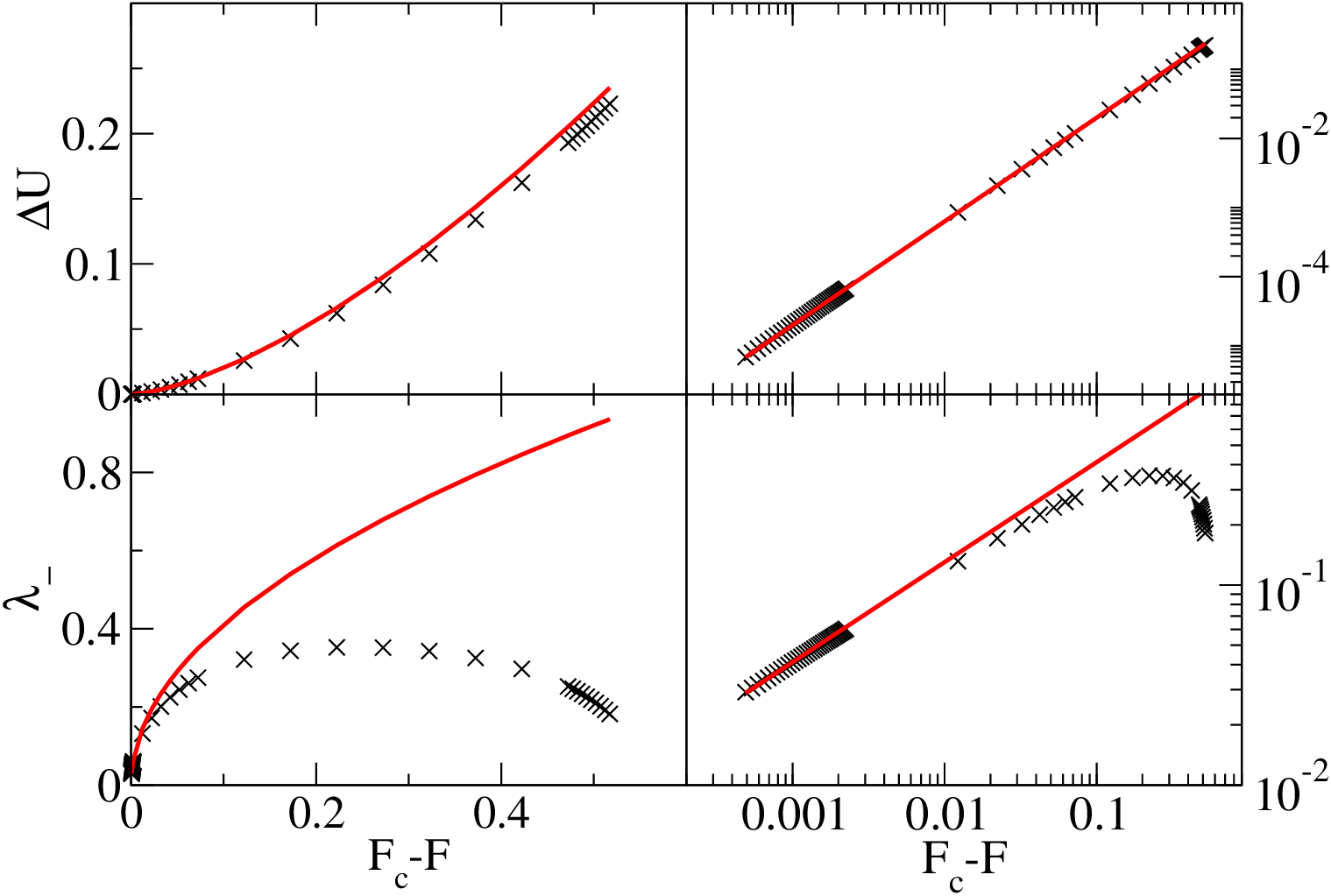}
  \caption{Top: $\Delta U$ and Bottom: $\lambda_-$ for the protein model, as functions of $F_c-F$ where $F$ is the external force and $F_c$ is the value of force at which the minimum and barrier collide.}
  \label{fig:proteinForce}
\end{figure}

To use force as a control parameter in the protein model, $L$ is reinstated as a \emph{bona fide} degree of freedom, and an external coupling is introduced so that $U_{\text{tot}}=U_{\text{int}}-FL$, where $U_{\text{int}}$ is the usual internal energy of the protein.
As the arguments for the fold scaling relations appeal only to the smoothness of the energy function and not the particular mode of loading, we expect analogous results when force is the control parameter.
$\Delta U$ and $\lambda_-$ are plotted in figure~\ref{fig:proteinForce}.
Again, all scaling relations are accurate at small $\delta$, but at large $\delta$ the scaling relation for the $\Delta U$ is much more accurate than the scaling relation for $\lambda_-$.

In summary, barrier heights in molecular systems are found to follow, to fairly high accuracy, the scaling relation $\Delta U\sim\delta^{3/2}$.  While this scaling relation has been known to be rigorously valid in the $\delta\rightarrow 0$ limit, this vanishing $\delta$ regime is not 
not physically significant at finite temperature, because thermal fluctuations cause the system to cross the barrier before the low delta regime is reached.  
However, our investigation shows that the scaling relation is appropriate outside of the low $\delta$ limit -- even when the scalings fail dramatically for the the curvatures or the fold ratio.  The barrier height scaling is relevant for all driven thermal systems, including flowing liquids, mechanically deformed glasses, and stretched proteins.  Quantitative analyses of these driven thermal systems, based on modifications of Eyring's theory to take the proper scalings of the barrier height into account, will lead to an improved understanding and description of these systems.

We thank W. Johnson,  M. Demkowicz, and A. Lema\^{i}tre for useful discussions.
CEM was supported under the auspices of the U.S. Department of Energy by the University of California, Lawrence Livermore National Laboratory under Contract No. W-7405-Eng-48.
DJL was supported by the National Science Foundation under Grant No. DMR-0402867

\end{document}